\documentclass[letterpaper,11pt]{article}
\usepackage{graphicx}
\usepackage{macros}
\usepackage{fullpage}
\usepackage{amsmath, amssymb, amsthm}

\newtheorem{theorem}{Theorem}
\newtheorem{lemma}{Lemma}

\newtheorem{remark}{Remark}

\newtheorem{observ}{Observation}

\begin{document}
	\title{Skyline Computation with Noisy Comparisons
	 }
	
	\author{Benoît Groz
	\thanks{Université Paris-Saclay, CNRS, LRI. {\tt benoit.groz@lri.fr}}
	\and 	Frederik Mallmann-Trenn		
	\thanks{King's College London. {\tt frederik.mallmann-trenn@kcl.ac.uk}}
	\and Claire Mathieu
	\thanks{CNRS \& IRIF. {\tt claire.m.mathieu@gmail.com}}
	\and Victor Verdugo
	\thanks{London School of Economics and Political Science. {\tt v.verdugo@lse.ac.uk}}
	\thanks{Universidad de O'Higgins. {\tt victor.verdugo@uoh.cl}}
	}
\date{\vspace{-1em}}
\maketitle
\begin{abstract}
Given a set of $n$ points in a $d$-dimensional space, we seek to compute the \emph{skyline}, i.e., those points that are not strictly dominated by any other point, using few comparisons between elements.
We adopt the noisy comparison model~\cite{FRPU94}  where comparisons fail with constant probability and confidence can be increased through independent repetitions
of a comparison.
In this model motivated by Crowdsourcing applications, Groz \& Milo \cite{GM15} show three bounds on the query complexity for the skyline problem. 
We improve significantly on that state of the art and provide two output-sensitive algorithms computing the skyline with respective query complexity \allowbreak
$O(nd\;\log (dk/\delta)) $ and $O(ndk\;\log (k/\delta)) $, where $k$ is the size of the skyline and $\delta$ the expected probability that our algorithm fails to return the correct answer. 
These results are tight for low dimensions. 
\end{abstract}
\newpage

\section{Introduction}\label{sec:intro}
\begin{figure}[H]
 \centering
 	\includegraphics[width=2in]{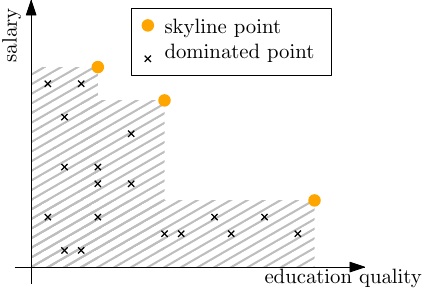}
 	\caption{ Given a set of points $X$, the goal is to find the set of \emph{skyline points}, i.e.,points are not dominated by any other points.  }\label{fig:ex} 
 \end{figure}

Skylines have been studied extensively, since the 1960s 
in statistics~\cite{Barndorff-Nielsen1966}, then in algorithms and computational geometry
 ~\cite{Kung:1975} and in databases~\cite{Borzsonyi:2001,Chomicki:2013,Godfrey:2007,Kossmann:2002}. Depending on the field of research, the {\em skyline} is also known as the set of {\em maximum vectors}, the {\em dominance frontier}, {\em admissible} points,
or {\em Pareto frontier.}
The  skyline of a set of points consists of those points which are not strictly dominated by any other point.
A point $p$ is \emph{dominated} by another point $q$ if $p_i\leq q_i$ for every coordinate (attribute or dimension) $i$. It is \emph{strictly dominated} if in addition the inequality is strict for at least one coordinate; see Figure~\ref{fig:ex}.
\vspace{.3cm}
 
\noindent{\it Noisy comparison model, and parameters.} 
In many contexts, comparing attributes is not straightforward.
Consider the example of finding {\em optimal} cities from \cite{GM15}.
\begin{quote} \em
	To compute the skyline with the help of the crowd we can ask people questions of the form ``is the education system superior in city x or city y?'' or ``can I expect a better salary in city x or city y''. Of course, people are likely to make mistakes, and so each question is typically posed to multiple people. Our objective is to minimize the number of questions that need to be issued to the crowd, while returning the correct skyline with high probability.
\end{quote}
Thus, much attention has recently been given to computing the skyline when information about the underlying data is uncertain~\cite{MWKMM11}, and comparisons may give erroneous answers. 
In this paper we investigate the complexity of computing skylines in the noisy comparison model, which was considered in~\cite{GM15} as a simplified model for crowd behaviour: we assume queries are  of the type {\it is the $i$-th coordinate of point $p$ (strictly) smaller than that of point $q$?},  and the outcome of each such query is independently correct with probability greater than some constant better than $1/2$ (for definiteness we assume probability $2/3$). As a consequence, our confidence on the relative order between $p$ and $q$ can be increased by repeatedly querying the pair on the same coordinate. Our complexity measure is the number of comparison queries performed.\\

This noisy comparison model was introduced in the seminal paper \cite{FRPU94} and has been studied in~\cite{GM15, BMW16}. There are at least two straightforward approaches to reduce noisy comparison problems to the noiseless comparison setting. One approach is to take any "noiseless" algorithm and repeat each of its comparisons $\log(f(n))$ times, where $n$ is the input size and $f(n)$ is the complexity of the algorithm. The other approach is to sort the $n$ items in all $d$ dimensions at a cost of $nd\log (nd)$, then run some noiseless algorithm based on the computed orders. The algorithms in~\cite{FRPU94,GM15} and this paper thus strive to avoid the logarithmic overhead of these straightforward approaches. \\

Three algorithms were proposed in~\cite{GM15} to compute skylines with noisy comparisons. Figure~\ref{fig:complexities} summarizes their complexity and the parameters we consider. 
The first algorithm  is the reduction through sorting discussed above. 
But skylines often contain only a small fraction of the input items (points), especially when there are few attributes to compare (low dimension). This leads to more efficient algorithms because smaller skylines are easier to compute.
Therefore, \cite{GM15} and the present paper expresses the complexity of computing skylines as a function of four parameters that appear on Figure~\ref{fig:complexities}: $\delta$, the probability that the algorithm could fail to return the correct output, and three parameters wholly determined by the input set $X$: the number of input points $n=|X|$, the dimension $d$ of those points, and the size $k=|\sky(X)|$ of the skyline (output). 
There is a substantial gap between the lower bounds and the upper bounds achieved by the skyline algorithms in~\cite{GM15}. In particular, the authors raised the question whether the skyline could be computed in $o(nk)$ for any constant $d$. In this paper, we tighten the gap between the lower and upper bounds and settle this open question.
\medskip

\noindent{\it Contributions.} 
We propose 2 new algorithms that compute skylines with probability at least $1-\delta$ and establish a lower bound:
\begin{itemize}
\item Algorithm $\guess(X,\delta)$ computes the skyline  in $O(nd\log(dk/\delta))$ query complexity and $O(nd\;\log(dk/\delta)+ndk)$ overall running time.
\item Algorithm $\guessb(X,\delta)$ computes the skyline  in $O(ndk\;\log(k/\delta))$
\item  $\Omega(nd\log k)$ queries are necessary to compute the skyline when $d\leq k$.
\item Additionally, we show that Algorithm $\guess$ can be adapted to compute the skyline with $O(nd\;\log(dk))$ comparisons  in the noiseless setting.
\end{itemize}
Our first algorithm answers positively the above question from~\cite{GM15}. Together with the lower bound, we thus settle the case of low dimensions, i.e., when there is a constant $c$ such that $d\leq k^c$. 
Our $2$ skyline algorithms both shave off a factor $k$ from the corresponding bounds in the state of the art~\cite{GM15}, as illustrated in Figure~\ref{fig:complexities} with respect to query complexity. 
$\guess$ is a randomized algorithm that samples the input, which means it may fail to compute the skyline within the bounds even when comparisons are guaranteed correct. However, we show that our algorithm can be adapted to achieve deterministic $O(nd\;\log (dk))$ for this specific noiseless case. 
As a subroutine for our algorithms, we developped a new algorithm to evaluate disjunctions of boolean variables with noise. Algorithm $\noisyft$ is, we believe, interesting in its own right: it returns the index of the first positive variable in input order, with a running time that scales linearly with the index. 

\begin{figure}[h]
\footnotesize
\begin{minipage}{.69\linewidth}
\renewcommand{\arraystretch}{1.5} 
\[\begin{array}{|l|c|c|c|}
\hline
\text{\cite{GM15}}& O(nd\;\log(nd/\delta))^\dagger &	O(ndk\;\log(dk/\delta))& 	O(ndk^2\log(k/\delta))\\\hline
{\text{this paper}}	&\rule[.1cm]{.5cm}{.01cm}&{O(nd\;\log(dk/\delta))}^\dagger& 	{O(ndk\;\log(k/\delta))}\\
&&\guess&\guessb\\\hline
\multicolumn{1}{l}{\color{Grey!70!black}\text{best when:}}&\multicolumn{1}{c}{\color{Grey!70!black}k\in\Omega(\log(dn))}	&\multicolumn{1}{c}{\color{Grey!70!black}d\leq k^c	\leq n}&\multicolumn{1}{c}{\color{Grey!70!black}k \ll d}\\
\end{array}
\]
\end{minipage}\qquad
\begin{minipage}{.2\linewidth}
\begin{tabular}{l<{\,:}@{\,}l}
$d$& dimension\\
$n$&  $\#$ input points\\
$k$& $\#$ skyline points\\
$\delta$& error rate tolerated
\end{tabular}
\end{minipage}
\caption{Query complexity of skyline algorithms depending on the values of $k$. 
For ${}^\dagger$-labeled bounds, the running time is larger than the number of queries.}\label{fig:complexities}
\end{figure}
\medskip

\noindent{\it Technical core of our algorithms.} 
The algorithm underlying the two bounds for $k\ll n$  in~\cite{GM15} recovers the skyline points one by one. It iteratively adds to the skyline the maximum point, in  lexicographic order, among those  not dominated by the  skyline points already found.
\footnote{The difference between those two bounds is due to different subroutines to check dominance. }
However, the algorithm in~\cite{GM15} essentially considers the whole input for each iteration. Our two algorithms, on the opposite, can identify and discard some dominated points early. 
The idea behind our algorithm $\algb$ is that it is more efficient to separate the two tasks: (i) finding a point $p$ not dominated by the  skyline points already found, on the one hand, and (ii) computing a maximum point (in lexicographic order) {\em among those dominating $p$}, on the other hand. 
Whenever a point is considered for step (i) but fails to satisfy that requirement, the point can be discarded definitively. 
The $O(ndk)$ skyline algorithm from~\cite{Clarkson94} for the noiseless setting also decomposes the two tasks, although the point they choose to add to the skyline in each of the $k$ iteration is not the same as ours.

Our algorithm $\alga$ can be viewed as a two-steps algorithm where the first step prunes a huge fraction of dominated points from the input through discretization, and the second step applies a cruder algorithm on the surviving points. We partition the input into buckets for discretization, identify ``skyline buckets'' and discard all points in dominated buckets.
The bucket boundaries are defined by sampling the input points and sorting all sample points in each dimension.
In the noisy comparison model, the approach of sampling the input for some kind of discretization was pioneered in~\cite{BMW16} for selection problems, but with rather different techniques and objectives.
One interesting aspect of our discretization is that a  fraction of the input will be, due to the low query complexity, incorrectly discretized yet we are able to recover the correct skyline. 

Our lower bound constructs a technical reduction from the problem of identifying null vectors among a collection of vectors, each having at most one non-zero coordinate. That problem can be studied using a two-phase process inspired from~\cite{FRPU94}.
\medskip

\noindent{\it Related work.} 
The noisy comparison model was considered for sorting and searching objects~\cite{FRPU94}. While any algorithm for that model can be reduced to the noiseless comparison model at the cost of a logarithmic factor (boosting each comparison so that by union bound all are correct), \cite{FRPU94} shows that this additional logarithmic factor can be spared for sorting and for maxima queries, though it cannot be spared for median selection. \cite{Newman09},\cite{GoyalS10} and~\cite{BMW16} investigate the trade-off between the total number of queries and the number of rounds for (variants of) top-k queries in the noisy comparison model and some other models. The noisy comparison model has been refined in~\cite{DavidsonKMR14} for top-k queries, where the probability of incorrect answers to a comparison increase with the distance between the two items. 

Other models for uncertain data have also been considered in the literature, where the location of  points is determined by a probability distribution, or when data is incomplete. Some previous work~\cite{Pei:2007,Afshani:2011} model uncertainty about the output by computing a $\rho$-skyline: points having probability at least $\rho$ to be in the skyline.
We refer to \cite{asudeh:2015} for skyline computation using the crowd and~\cite{LiWZF17} for a survey in crowdsourced data management.\\

Our paper aims to establish the worst-case number of comparisons required to compute skylines with output-sensitive algorithms, i.e., when the cost is parametrized by the size of the result set. While one of our algorithm is randomized, we do not make any further assumption on the input (we do not assume input points are uniformly distributed, for instance). 
In the classic {\it noiseless} comparison model, the problem of computing skylines has received a large amount of 
attention~\cite{Kung:1975,Borzsonyi:2001,KS85}.  
For any constant $d$, \cite{KS85} show that skylines can be computed in $O(n\log^{d-2}k)$. In the RAM model, the fastest algorithms we are aware of run in $O(n\log^{d-3}n)$ expected time~\cite{ChanLP11}, and $O(n\log\log_{n/k}n (\log_{n/k}n)^{d-3}$ deterministic time~\cite{Afshani14}. 
When $d\in \{2,3\}$, the problem even admits ``instance-optimal" algorithms~\cite{Afshani3046673}.
\cite{ChanL15} investigates the constant factor for the number of comparisons required to compute skyline, when $d\in \{2,3\}$. The technique does not seem to generalize to arbitrary dimensions, and the authors ask among open problems whether arbitrary skylines can be computed with fewer than $dn\log n$ comparisons. To the best of our knowledge, our $O(nd\log(dk))$ is the first non-trivial output-sensitive upper bound that improves on the folklore $O(dnk)$ for computing skylines in arbitrary dimensions. 
Many other algorithms have been proposed that fit particular settings (big data environment, particular distributions, etc), as evidenced in the survey~\cite{KalyvasT17}, but those works are further from ours as they generally do not investigate the asymptotic number of comparisons.
Other skyline algorithms in the literature for the noiseless setting have used bucketing. In particular,~\cite{AfratiKSU12} computes the skyline in a massively parallel setting by partitioning the input based on quantiles along each dimension. This means they define similar buckets to ours, 
and they already observed that the buckets that contain skyline points are located in hyperplanes around the "bucket skyline", and therefore
those buckets only contain a small fraction of the whole input.\\

\noindent{\it Organization. }In Section~\ref{sec:preliminaries}, we recall standard results about the noisy comparison model and introduce some procedure at the core of our algorithms.
Section~\ref{sec:highdim} introduces our algorithm for high dimensions (Theorem~\ref{thm:highdim}) and Section~\ref{sec:lowdim} introduces the counterpart for low dimensions (Theorem~\ref{thm-lowdim}).
Section~\ref{sec:LB} establishes our lower bound (Theorem~\ref{THM:LOWER}).

\section{Preliminaries}\label{sec:preliminaries}
The complexity measured is the number of comparisons in the worst case. Whenever the running time and the number of comparisons differ, we will say so. With respect to the probability of error, our algorithms are supposed to fail with probability at most $\delta$. Following standard practice we only care to prove that our algorithms have error in $O(\delta)$: $5\delta$, for instance, because the asymptotic complexity of our algorithms would remain the same with an adjusted value for the parameter: $\delta'=\delta/5$.
Given two points, $p=(p_1,p_2\dots, p_d)$ and $q=(q_1,q_2\dots, q_d)$, we say that
point $p$ is {\it lexicographically} smaller than $q$, denoted by $p \leq_{\text{lex}}q$ , if $p_i<q_i$ for the first $i$ where $p_i$ and $q_i$ differ. 
If there is no such $i$, meaning that the points are identical, we use the id of the points in the input as a tie-breaker, ensuring that we obtain a total order.
We next describe and name algorithms that we use as subroutines to compute skylines.\\

\noindent Algorithm $\noisysearch$ takes as  input 
an element $y$, 
an ordered list $(y_1,y_2,\ldots ,y_{m})$, accessible by comparisons that each have error probability at most $p$, 
and a parameter $\delta$. 
The goal is to output the interval $I=(y_{i-1},y_i]$ such that $y\in I$.\\

\noindent Algorithm $\noisysort$ relies on $\noisysearch$ to solve the {\bf noisy sort problem}. 
It takes as input 
an unordered set $Y=\{ y_1,y_2,\ldots ,y_{m}\}$, 
and a parameter $\delta$. 
The goal is to output an ordering of $Y$ that is the correct non-decreasing sorted order. 
In the definition above, the order is kept implicit. In our algorithms, the input items are $d$-dimensional points, so $\noisysort$ will take an additional argument $i$ indicating on which coordinate we are sorting those points. \\

\noindent Algorithm $\noisymax$ returns the maximum item in the unordered set $Y$ whose elements can be compared, but we will rather use another variant:
algorithm $\maxlex$ takes as input 
an unordered set $Y=\{ y_1,y_2,\ldots ,y_{m}\}$, 
a point $x$ 
and a parameter $\delta$. 
The goal is to output the maximum point in lexicographic order among those that dominate $x$. 
Algorithm $\setdominates$ is the boolean version whose goal is to output whether there exists a point in $Y$ that dominates $x$.\\

\noindent Algorithm $\noisyft$ takes as input 
a list  $( y_1,y_2,\ldots ,y_{m})$ of boolean elements  that can be compared to \true with error probability at most $p$
(typically the result of some comparison or subroutines such as $\setdominates$).
The goal is to output the index of the first  element with value \true (and $m+1$, which we assimilate to \false, if there are none). \footnote{As in~\cite{GM15} (but with stronger bounds), this improves upon an $O(m\log(1/\delta))$ algorithm from~\cite{FRPU94} that only answers whether at least one of the elements is true.}

\begin{theorem}[\cite{FRPU94},\cite{GM15}]\label{thm:noisy-search}
When the input comparisons have error probability at most $p=1/3$, 
the table below lists the number of comparisons performed by the algorithms to return the correct answer with success probability $1-\delta$:

\begin{table}[H]
\renewcommand{\arraystretch}{1.25}
\begin{tabular}{l||l|l|l|l|l}
 Algorithm& $\noisymax$ & $\noisysort $& $\noisysearch$  &$ \setdominates $& $\maxlex $\\
 \hline
 Comparisons & $O(m\log\frac{1}{\delta})$ &
  $O(m\log\frac{m}{\delta})$ & 
  $ O(\log \frac{m}{\delta})$&
   $O(md\;\log\frac{1}{\delta})$ & $O(md\;\log\frac{1}{\delta})$
\end{tabular}
\end{table}
\end{theorem}

We denote by $\text{CheckVar}(x,\delta)$ the procedure that checks if $x=\true$ 
with error probability $\delta$ by majority vote, and returns the corresponding boolean. 
\begin{theorem}
\label{thm:first-positive}
Algorithm $\noisyft$ solves the first positive variable problem with success probability $1-\delta$ in $O(j\; \log(1/\delta))$ where $j$ is the index returned.
\end{theorem}
\begin{proof}
The proof, left for the appendix~\cite{abs-1710-02058}, shows that the error (resp. the cost) of the whole algorithm is dominated by the error (resp. the cost) of the last iteration.
\end{proof}

\begin{algorithm*}\caption*{$\operatorname{\bf Algorithm}$\ $\noisyft(x_1,\dots,x_n,\delta)$     \  \ \ \ \ (see Theorem~\ref{thm:first-positive})
\\
{\bf input}: $\{x_1, \dots, x_n\}$ set of boolean random variables, $\delta$ error probability\\
{\bf output}:  the index $j$ of the first positive variable, or $n+1$ (=\;\false).
}
\label{alg:or-trustpreserv-outsens}
\begin{algorithmic}[1]
\STATE $i\gets 1$
\STATE $\delta'\gets \delta/2$
\WHILE{$i\leq n$}
\STATE $j\gets \noisyor(x_1,\dots,x_i,\delta')$
\IF{$\text{CheckVar}(x_j,\delta'/2^i)$}
\RETURN  $j$
\ELSE 
\STATE $i\gets 2\cdot i$
\ENDIF
\ENDWHILE
\RETURN \false
\end{algorithmic}
\end{algorithm*}

\section{Skyline Computation in High Dimension}\label{sec:highdim}
We first introduce Algorithm~$\algb$ which assumes that an estimate $\hat{k}$ of $k$ is known in advance. We will show afterwards how we can lift that assumption.

\begin{algorithm*}\caption*{$\operatorname{\bf Algorithm}$\ $\algb(k,X,\delta)$     \  \ \ \ \ (see Theorem~\ref{thm:highdim-kknown})
\\
{\bf input}: $X=\{p_1, \dots, p_n\}$ set of points, $\hat{k}$ upper bound on skyline size, $\delta$ error probability\\
{\bf output}:    $\min(\hat{k},\sky(X))$ skyline points w.p. $1-\delta$
}
\begin{algorithmic}[1]
\label{alg:skyline-low-dim}
\STATE Initialize $S\gets \emptyset$, $i\gets 1$
\WHILE{ $i\neq -1$ and $|S|<\hat{k}$}
\STATE $i'\gets $ index of  the first point $p_{i'}$ not dominated by current skyline points.\footnotemark{}\\[.1cm]
\COMMENT {{\color{darkgreen} Find a skyline point dominating $p_i$}}
\STATE Compute  $ p^* \leftarrow \maxlex(p_{i'},\{p_i,\dots, p_n\}, \delta/(2\hat{k}))$ \label{line:maxlex}
\STATE{$S\gets S\cup \{ p^*\}$}
\STATE{$i\gets i'$}
\ENDWHILE
\STATE Output {$S$}
\end{algorithmic}
\end{algorithm*}
\footnotetext{This can be computed using algorithm $\noisyft$ on the boolean variables: $\neg\setdominates(S,p_i,\delta/(2\hat{k}))$,\dots, $\neg\setdominates(S,p_n,\delta/(2\hat{k}))$,
 where we denote by $\neg$ the negation. This means that $\neg\setdominates(S,p_n,\delta/(2\hat{k}))$ returns true when the procedure $\setdominates(S,p_n,\delta/(2\hat{k}))$ indicates that $p_n$ is not dominated.
}
\begin{theorem}\label{thm:highdim-kknown}
Given $\delta \in (0,1/2)$ and a set $X$ of data items,  
$\algb(X,\delta)$ outputs $\min(|X|,\hat{k})$ skyline points, with probability at least $1-\delta$. 
The  running time and number of queries is $O(nd\hat{k}\log(\hat{k}/\delta)).$
\end{theorem}
\begin{proof}
Each iteration through the loop adds a point to the skyline $S$ with probability of error at most $\delta/\hat{k}$. The final result is therefore correct with success probability $1-\delta$.  
The complexity is $O((i'-i)\cdot d\hat{k}\log(\hat{k}/\delta))$ to find a non-dominated point $p_{i'}$ at line $3$, and $O(nd\;\log(\hat{k}/\delta))$ to compute the maximal point above $p_{i'}$ at line $4$.
Summing over all iterations, the running time and number of queries is $O(nd\;\hat{k}\log(\hat{k}/\delta))$. 
\end{proof}

Algorithm $\algb(X,\delta)$ needs a good estimate of the skyline cardinality $\hat{k}\in O(k)$ to  return the skyline in $O(ndk\;\log(k/\delta))$.
To guarantee  that complexity, algorithm~$\guessb$ exploits the classical trick from Chan~\cite{Chan96} of trying  a sequence of successive values for $\hat{k}$ -- a trick that we also exploit in algorithms~$\noisyft$ and~$\guess$. The  sequence grows exponentially to prevent failed attempts from penalizing the complexity. 

\begin{algorithm}[htb]
\caption*{$\operatorname{\bf Algorithm}$\ $\guessb(X,\delta )$     \  \ \ \ \ (see Theorem~\ref{thm:highdim})
\\
{\bf input:} $X$ set of points, $\delta$ error probability \\
{\bf output:} $\sky(X)$ w.p. $1-\delta$
}
\begin{algorithmic}[1]
\label{alg:skyline2} 
\STATE Initialize $j\gets 0$, $\hat{k}\gets 1$ 
\REPEAT 
	\STATE $j\gets j+1$ ; $\hat{k}\gets 2\hat{k}$ ; $S\gets\algb(\hat{k},X,\delta/2^j)$
\UNTIL { $|S| < \hat{k}$}
\STATE Output $S$
\end{algorithmic}
\end{algorithm}

\begin{theorem}\label{thm:highdim}
Given $\delta \in (0,1/2)$ and a set $X$ of data items,  
$\guessb(X,\delta)$ outputs a subset of $X$ which, with probability at least $1-\delta$, is the skyline. 
The  running time and number of queries is $O(ndk\;\log(k/\delta)).$
\end{theorem}
\begin{proof}
The proof is relatively straightforward and left for the appendix.
\end{proof}

\section{Skyline Computation in Low Dimension}\label{sec:lowdim}

Let us first sketch our algorithm $\alga(k,X,\delta)$. 
The algorithm works in three phases. The first phase partitions input points in buckets.  We sort the $i$-th coordinate of a random sample to define $s+1$ intervals in each dimension $i\in [d]$, hence $(s+1)^d$ {\em buckets}, where each bucket is a product of intervals of the form $\prod_i I_i$; then  we assign each  point $p$ of $X$ to a  bucket by searching in each dimension for the interval $I_i$ containing $p_i$. Of course we do not materialize buckets that are not assigned any points.

The second phase eliminates irrelevant buckets: those that are dominated by some non-empty bucket and therefore have no chance of containing a skyline point. In short, the idea is to identify the "skyline of the buckets", and use it to discard the dominated buckets, as defined in  Section~\ref{subsec:domination-buckets}.
With high probability the bucketization obtained from the first phase will be "accurate enough" for our purpose: it will allow  to identify efficiently the irrelevant buckets, and will also guarantee that the points in the remaining buckets form a small fraction of the input (provided $k$ and $d$ are small). 

Finally, we solve the skyline problem on a much smaller dataset, calling Algorithm $\algb$ to find the skyline of the remaining points.
\footnote{Alternatively, one could use an algorithm provided by Groz and Milo~\cite{GM15},  it is only important that the size of the input set is reduced to $n/k$ to cope with the larger runtime of the mentioned algorithms.} 
 The whole purpose of the bucketization is  to discard most input points while preserving the actual skyline points, so that we can then run a more expensive algorithm on the reduced dataset.
 

\subsection{Identifying "Truly Non-empty" Buckets}

Our bucketization does not guarantee that all points are assigned to the proper bucket, because it would be too costly with noisy comparisons.
In particular, empty buckets may erroneously be assumed to contain some points (e.g., the buckets above $a,b$ on Figure~\ref{fig:ex2}). Those empty buckets also are irrelevant, even if they are not dominated by the "skyline" buckets.
To drop the irrelevant buckets, we thus design a subroutine {\bf First-Nonempty-Bucket} that processes a list of buckets, and returns the first bucket that really contains at least one point. Incidentally, we will not double-check the emptiness of every bucket using this procedure, but will only check those that may possibly belong to the skyline: those that we will define more formally as buckets of type (i), (ii) and (iv) in the proof of Theorem~\ref{thm-lowdim-kknown}. 
We could not afford to "fix" the whole assignment as it may contain too many buckets.

In the {\bf First-Nonempty-Bucket} problem, the input is a sequence of pairs $[(B_1,X_1),\dots,(B_n,X_n)]$ where $B_i$ is a bucket and $X_i$ is a set of points. 
The goal is to return the first $i$ such that $B_i\cap X_i\neq\emptyset$ with success probability $1-\delta$. The test $B_i\cap X_i\neq \emptyset$ can be formulated as a DNF with $|X_i|$ conjunctions of $O(d)$ boolean variables each. To solve {\bf First-Nonempty-Bucket}, we can flatten the formulas of all buckets into a large DNF with conjunctions of $O(d)$ boolean variables (one conjunction per bucket point). We call $\noisyfb([(B_1,X_1),\dots,(B_n,X_n)],\delta)$ the algorithm that executes $\noisyft$ to compute the first true conjunction, while keeping tracks of which point belongs to which bucket with pointers:

\begin{lemma}\label{lem:emptiness-test}
	Algorithm $\noisyfb([(B_1,X_1),\dots,(B_n,X_n)],\delta)$ solves problem {\bf First-Nonempty-Bucket}
	 in $O(d\;\sum_{i\leq j} |X_i|\log(1/\delta))$ with success probability $1-\delta$, where $j$ is the index returned by the algorithm.
\end{lemma}

\subsection{Domination Relationships Between Buckets}\label{subsec:domination-buckets}
In the second phase, Algorithm $\alga(k,X,\delta)$ eliminates irrelevant buckets. 
To manage ties, we need to distinguish two kinds of intervals: the trivial intervals that match a sample coordinate: 
$I=[x,x]$, and the non-trivial intervals $I=]a,b[$ ($a<b$) contained between samples (or above the largest sample, or below the smallest sample). 
To compare easily those intervals, we adopt the convention that for a non-trivial interval $I=]a,b[$, 
$\min\ I=a+\epsilon$ and $\max\ I=b-\epsilon$ for some infinitesimal $\epsilon>0$; $\epsilon=(b-a)/3$ would do.
We say that a bucket $B=\prod_i I_i$ is \emph{dominated} by a different bucket $B'=\prod_iI'_i$ if in every dimension $\max\ I_i\leq \min\ I'_i$. 
Equivalently: we say that $B'$ dominates $B$ if every point (whether in the dataset or hypothetical) in $B'$ dominates every point in $B$. 
The idea is that no skyline point belongs to a bucket dominated by a non-empty bucket. 
We observe that the relative position of buckets is known by construction, 
so deciding whether a bucket dominates another one may require time $O(d)$ but does not require any comparison query. \\

\begin{figure}[H]
\centering
	\includegraphics[width=5in]{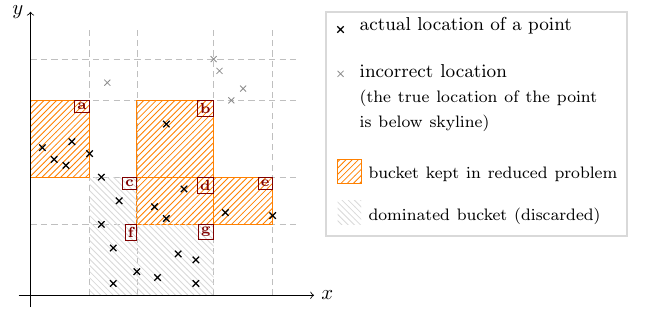}
	\caption{ An illustration of the bucket dominance and its role in $\alga$. 
	Here bucket $b$ dominates $c$ and $f$ but not $a$, $d$, $e$ or $g$.
	Buckets $c,f,g$ are dominated by some non-empty bucket and therefore cannot contain a skyline point. Bucket $a$ does not contain a skyline point, but this cannot be deduced from the bucket assignments, therefore points in bucket $a$ are passed on to the reduced problem. In this figure we may assume to simplify that a bucket contains its upper boundary. But in our algorithm bucket $a$ would actually contain only the four leftmost points, and the fifth point would belong to a distinct bucket with a trivial interval on $x$.}
	\label{fig:ex2} 
\end{figure}

Figure~\ref{fig:ex2} illustrates the relevant and discarded buckets. 
On that figure, we depicted a few empty buckets above the skyline that are erroneously assumed to contain some points as a result of noise during the assignment. Of course, there are also incorrect assignments of points into empty or non-empty buckets below the skyline, as well as incorrect assignments into the "skyline buckets". These incorrect assignments are not an issue as long as there are not too many of them: dominated buckets will be discarded as such, whether empty or not, and the few irrelevant points maintained into the reduced dataset will be discarded in phase~3, when the skyline of this dataset is computed.

\subsection{Algorithm and Bounds for Skyline Computation in Low Dimension}
\begin{algorithm}
\caption*{$\operatorname{\bf Algorithm}$\ $\alga(\hat{k},X,\delta)$ \  \ \ \ \ (see Theorem~\ref{thm-lowdim-kknown})
\\
{\bf input}:  $\hat{k}$ integer, $X$ set of points, $\delta$ error probability\\
{\bf output}: $\min\{\hat{k},|\sky(X)|\}$ points of $\sky(X)$\\
{\bf error probability}: $\delta$
 }
\begin{algorithmic}[1]
\label{alg1-kknown}
\IF{ $\hat{k}^5\geq n$ or $d^5\geq n$ or $(\log (1/\delta))^5\geq n$} \label{line:kdbig}
\STATE Compute the skyline by sorting every dimension, as in~\cite{GM15}. Return that skyline.
\ENDIF
\STATE $\delta' \gets \delta/(2d\hat{k})^5$  and $s\gets d\hat{k}^2\log (d^2 \hat{k}^2/\delta')$\label{line:algareducedelta}\\[.1cm]
\COMMENT{{\color{darkgreen} Phase (i): bucketing}}
\FOR{each dimension $i\in \{1,2,\ldots,d\}$}
	\STATE $S_i\gets \noisysort($sample of $X$ of size $s,i,\delta'/d)$ \label{line:sortsample}
	\STATE Remove duplicates so that, with prob. $1-\delta'/d$, the values in $S_i$ are all distinct.\footnotemark  \label{line:mergeduplicates}\\
\ENDFOR
\FOR{each point $p\in X$}
	\STATE Place $p$ in set $X_B$ associated to $B=\prod_{i=1}^dI_i$, with $I_i= \noisysearch (p_i,S_i,{\delta'}/{(d\hat{k})})$. \\
	 \label{line:bucket}
\ENDFOR
\STATE Drop all empty buckets (those that were assigned no point).\label{line:dropempty}
\STATE Sort buckets into a sequence $B_1,\dots, B_h$ so that each bucket comes before buckets it dominates.\label{line:sortbuckets}\\[.1cm]
\COMMENT{{\color{darkgreen} Phase (ii): eliminating irrelevant buckets}}
\STATE Initialize $X'\gets \emptyset$, $i\gets 1$
\WHILE{ $i\neq -1$}
\STATE $i\gets \noisyfb([(B_1,X_{B_1}),\dots,(B_h,X_{B_h})],\delta'/\hat{k}))$\label{line:empty}
\STATE $X'\gets X'\cup X_{B_{i}}$
\IF{$|X'| > 8n/\hat{k}$}\label{line:check-cardinality-survivors}
\STATE Raise an error.
\ENDIF
\STATE Drop from $B_{1},\dots, B_h$ all buckets dominated by $B_{i}$, and also buckets $B_1$ to $B_i$.\label{line:dropdominated}
\ENDWHILE

\vspace*{.1cm}\COMMENT{{\color{darkgreen} Phase (iii): solve reduced problem}}
\STATE Output $\guessb(X',\delta')$.\label{line:GM}
\end{algorithmic}
\end{algorithm}
\footnotetext{Note that $X$ can contain points sharing the same coordinate meaning  that the $S_i$ are not necessarily distinct. }

\begin{theorem}\label{thm-lowdim-kknown}
Given $\delta \in (0,1/2)$, $\hat{k}>0$, and a set $X$ of data items,  algorithm
$\alga(\hat{k},X,\delta)$ outputs $\min(|X|,\hat{k})$ skyline points, with probability at least $1-\delta$. The  
number of queries is $O(nd\;\log(d\hat{k}/\delta))$. The running time is $O(nd\;\log(d\hat{k}/\delta)+nd\cdot\min(\hat{k},|\sky(X)|))$
\end{theorem}
\begin{proof}
The proof, left for the appendix, first  shows by Chernoff bounds that the assignment satisfies with high probability some key properties: (1) few points are erroneously assigned to incorrect buckets (2) the skyline points are assigned to the correct bucket, and (3) there are at most $O(n/(d\hat{k}^2))$ points on any hyperplane (i.e., in buckets that are ties on some dimension). 
The proof then shows that:
\begin{itemize}
\item There are at most  $O(n/\hat{k})$ points in the reduced problem. This is because those points belong to skyline buckets or buckets that are tied with a skyline bucket on at least one dimension (every other non-empty bucket is dominated), and property (3) of the assignment guarantees that the union of all such buckets has at most $O(n/\hat{k})$ points.
\item The buckets above the skyline buckets  which are erroneously assumed to contain points can quickly be identified and eliminated since they contain few points.\qedhere
\end{itemize}
\end{proof}

Algorithm $\alga(\hat{k},X,\delta)$ needs a good estimate of the skyline cardinality to return the skyline in $O(nd\;\log(dk/\delta))$: we must have $\hat{k}\geq k$ and $\log(\hat{k})\in O(\log(k))$. 
Algorithm $\guess(X,\delta )$  (left for the appendix) guarantees the complexity by trying a sequence of successive values for $\hat{k}$. The successive values in the sequence grow super exponentially (similarly to \cite{Chan96,GM15}) to prevent failed attempts from penalizing the complexity.

\begin{theorem}\label{thm-lowdim}
Given $\delta \in (0,1/2)$ and a set $X$ of data items,  
$\guess(\hat{k},X,\delta)$ outputs a subset of $X$ which, with probability at least $1-\delta$, is the skyline. 
The  number of queries is $O(nd\;\log(dk/\delta))$. The running time is $O(nd\;\log(dk/\delta)+ndk)$.
\end{theorem}
\begin{proof}
For iteration $j$, the algorithm bounds the probability of error by $\delta/2^j$, and the corresponding cost is given by Theorem~\ref{thm-lowdim-kknown}, hence the complexity we claim by summing those terms over all iterations.
\end{proof}

\begin{remark}
In the noiseless setting, we could adopt the same sampling approach to assign points to buckets and reduce the input size.
On line~\ref{line:GM} we could use any noiseless skyline algorithm such as the $O(ndk)$ algorithm from~\cite{Clarkson94}, or our own similar $\guessb$ which can clearly run in $O(ndk)$ in the noiseless case. 
The cost of the bucketing phase remains $O(nd\;\log (d\hat{k}/\delta))$.
The elimination phase becomes rather trivial since all points get assigned to their proper bucket, and therefore there is no need to check buckets for emptiness as in Line~\ref{line:empty}.
By setting $\delta=1/k$ failures are scarce enough so that the higher cost of $O(ndk)$ in case of failure is covered by the cost of an execution corresponding to a satisfying sample.
Consequently, the expected query complexity is $O(nd\;\log(dk))$, and the running time $O(nd\;\log (dk)+ndk)$.
Better yet: we can replace random sampling with quantile selection to obtain a deterministic algorithm with the same bounds. Algorithms for the \emph{multiple selection} problem are surveyed in~\cite{ChanL15}. Actually, our algorithm can be viewed as some kind of generalization to higher dimensions of an algorithm from~\cite{ChanL15}
which assigns points to buckets before recursing, the buckets being the quantiles along one coordinate.
\end{remark}

\section{Lower Bound}\label{sec:LB}
\newcommand{\A}{\mathcal{A}}

To achieve meaningful lower bounds (that do not reduce to the noiseless setting), we assume here that the input comparisons have a probability of error at least 1/3. Of course, we just need the probability to be bounded away from zero.
The proof of the following theorem is left for the appendix
\begin{theorem}\label{THM:LOWER}
For any $n\geq k\geq d>0$, any algorithm that recovers  with error probability at most $1/10$ the skyline for any input having exactly $k$ skyline points, requires $\Omega(nd\;\log (k/\delta))$ queries in expectation on a worst-case input. 
\end{theorem}

\section{Conclusion and Related Work}
We introduced two algorithms to compute skylines with noisy comparisons. The most involved shows that we can compute skylines in $O(nd\;\log (dk/\delta))$ comparisons. We also show that this bound is optimal when the dimensions is low ($d\leq k^c$ for some constant $c$), since computing noisy skylines requires $\Omega(dn\log k)$ comparisons.
All our algorithms but $\guess$ in $O(nd\;\log dk/\delta)$ are what we call \emph{trust-preserving}(\cite{GM15}), meaning that when the probability of errors in input comparisons is already at most $\delta<1/3$,  we can discard 
from the complexity the dependency in $\delta$ (replacing $\delta$ by some constant).

We leave open the question of the optimal number of comparisons required to compute skylines for arbitrarily large dimensions. Even in the noiseless case, it is not lear whether the skyline could be computed in $O(nd\;\log k)$ comparisons. 
Our algorithm is output sensitive (the running time is optimized with respect to the output size) but we did not investigate its instance optimality. However, knowing the input set up to a permutation of the points does not seem to help identifying the skyline points in the noisy comparison model, so we believe that for every $k$ and on any input of skyline cardinality $k$, even with this knowledge any skyline algorithm would still require $\Omega(nd\;\log k)$ comparisons. We leave open the question of establishing such a stronger lower bound.

\bibliography{biblio}
\bibliographystyle{splncs04}
\newpage

\section*{Appendix: Upper Bounds}

\begin{proof}[Proof of Theorem~\ref{thm:first-positive}]
We denote by $j'$ the true index of the first positive variable in $x_1,\dots, x_m$. 
We assume the input comparison have error probability at most $\delta$, which can be achieved at a cost by repeating queries $\log(1/\delta)$ times.
The probability that $\noisyor$ fails to identify $j'$ for $i=2^{\lceil \log k\rceil}$ (i.e., the first time it faces variable $x_{j'}$) is at most $\delta'$.
The probability that an incorrect index is returned (before $i\geq j'$) is at most $\sum_i \delta'/2^i$. The algorithm thus returns an incorrect index with probability at most $\delta'+\sum_i \delta'/2^i\leq\delta$. 
$\noisyor$ requires $O(i)$ comparisons at line~4, whereas $\text{CheckVar}$ requires $O(i)$ comparisons at line~5.
Replacing $i$ with $2^h$, the total cost on a successful execution is therefore $\sum_{h=1}^{\lceil \log (j')\rceil} 2^h=O(j')$.
\end{proof}

\begin{proof}[Proof of Theorem~\ref{thm:highdim}]
For iteration $j$, the probability of error is $\delta/2^j$, and the cost is $O(nd\hat{k}\;\log(\hat{k}/\delta))$.
Consequently, the probability that the algorithm fails to return the correct answer is at most
$\sum_j\delta/2^j\leq \delta$, and the running time is 
\[O\left(\sum_{j=1}^{\lfloor \log k\rfloor +1} nd2^j\;\log(2^j\cdot 2^j/\delta)\right)\in O(ndk\;\log(k/\delta)).\]
The complexity is $O((i'-i)\cdot d\hat{k}\;\log(\hat{k}/\delta))$ to find a non-dominated point $p_{i'}$ at line $4$, and $O(nd\;\log(\hat{k}/\delta))$ to compute the maximal point above $p_{i'}$ at line $6$.
Summing over all iterations, the running time and number of queries is $O(nd\hat{k}\;\log(\hat{k}/\delta))$. 
\end{proof}

\subsection*{Proof of Theorem~\ref{thm-lowdim-kknown}}

The following Lemma lists properties that our bucketing assignment satisfies with high probability. We will show in Theorem~\ref{thm-lowdim-kknown} that our algorithm can compute the skyline efficiently for any assignment satisfying those properties.
\begin{lemma}\label{lem:assignment-properties}
Assume that the samples have been correctly ordered
at line~\ref{line:sortsample}.
With error probability $\delta/\hat{k}$, the assignment performed at line~8 satisfies  the following two properties:
\begin{enumerate}
\item If $I$ is a non-trivial interval (i.e., unless it matches the coordinate of a sample point), 
\[|\{  p \colon I=\noisysearch(p_j,S_j,\delta'/(d\hat{k}))\}| \le 4n/(d\hat{k}^2).\]
\item Less than $2n/(d\hat{k}^2)$ points are (erroneously) assigned to buckets above the real skyline buckets. 
\item The skyline points are assigned to their correct bucket.
\end{enumerate}
\end{lemma}
\begin{proof}
Recall that $\delta'=\delta/(2d\hat{k})^5$, and that $p_j$ denotes the $j\textsuperscript{th}$ coordinate of point $j$.
Assume the points of $X$ are ordered w.r.t. to their $j$\textsuperscript{th} coordinate, breaking ties arbitrarily.
Consider these ordered points to be divided into blocks, each one having $\ell=n/(d\hat{k}^2)$ consecutive points, except the last which may have less.  
In particular, the number of blocks is $d\hat{k}^2$.

Consider now the samples after line~\ref{line:sortsample}. 
Each block (but the last) contains at least one sample  with probability at least $1-(1-\ell/n)^{s} \geq 1-\delta'/d^2\hat{k}^2$.
\bnote{I did not see why $s-1$ so I replaced with $s$}
If one sample is indeed taken from every block (except maybe the last), the distance between any two samples 
is at most $2\ell$. As a consequence, the number of points $p$ that should be assigned to any given bucket is bounded by $2\ell$, except for buckets with a trivial interval because several such buckets can be merged when removing duplicates at line~\ref{line:mergeduplicates}. 
By Chernoff bounds, the number of points assigned to wrong buckets is at most $2n/(d\hat{k}^2)$ w.p. at least $1-\delta'$.

By union bound over all $d$ dimensions and over all $s$ intervals, we therefore have probability at least $1-3\delta'$ that one sample is taken from each block and that the total number $w$ of points assigned to wrong buckets (over all dimensions and blocks) is less than $2n/(d\hat{k}^2)$.
Consequently, with probability at least $1-3\delta'$ the assignment satisfies the first property.
Indeed, for each dimension $j$ and interval $I$, the number of points in $I$ is bounded by 
$2\ell$ (maximum distance between two samples) plus $2n/(d\hat{k}^2)$ (incorrect assignments into buckets):
\[|\{  p \colon p \text{  was sorted into $I$ in line~\ref{line:bucket}}\}| \leq 2\ell+\frac{2n}{d\hat{k}^2}= \frac{4n}{d\hat{k}^2}.\]
As for the number of buckets erroneously assumed to be non-empty, it is bounded by the number of points assigned to wrong buckets and is therefore at most $2n/(d\hat{k}^2)$.
\end{proof}

This concludes the proof of the Lemma. 
We next turn to the proof of Theorem~\ref{thm-lowdim-kknown}.

\begin{proof}[Proof of Theorem~\ref{thm-lowdim-kknown}]
When $\hat{k}^5\geq n$, $d^5\geq n$ or $(\log (1/\delta))^5\geq n$, the bounds can clearly be achieved by the other algorithms discussed previously, so we assume w.l.o.g. that $\hat{k}^5<n$ and $d^5<n$ and $(\log (1/\delta))^5< n$.
We evaluate the cost of the algorithm assuming that (a) the samples are correctly sorted at Line~\ref{line:sortsample}, (b) the assignment satisfies the properties in Lemma~\ref{lem:assignment-properties}, and (c) no mistakes are made at lines~\ref{line:empty} and~\ref{line:GM}. In other words, we only accept a few mistakes at Line~\ref{line:bucket}.\\

\noindent {\bf Phase (i) Bucketing.} 
 Line~\ref{line:sortsample}: by Theorem~\ref{thm:noisy-search} (noisy sorting) the sample is sorted in
 $d\cdot O(s\log(sd/\delta'))=O(nd\log(d\hat{k}/\delta))$.
Line~\ref{line:bucket}: by Theorem~\ref{thm:noisy-search} (noisy search) the points are assigned to their bucket in
 $nd\cdot O(\log(sd\hat{k}/\delta'))=O(nd\;\log (d\hat{k}/\delta))$.
We will distinguish four kinds of (presumably) non-empty buckets (all other buckets are dropped at line~\ref{line:dropempty}):  
(i) those above the skyline that have been erroneously assigned some points, 
(ii) the buckets containing skyline points, 
(iii) the buckets that are dominated by buckets of type (ii), 
and (iv) the other (non-empty) buckets: they are not above the skyline but we do not have sufficient information to realize that they have no skyline points, because they are not dominated by any non-empty bucket. 
The algorithm is obviously not able to distinguish buckets of type (ii) and (iv), hence both are passed on to $\algb$ at line~\ref{line:GM}.
 
The number $h$ of non-empty buckets is not necessarily much smaller than $n$ as $h$ may grow exponentially with $d$.
Line~\ref{line:sortbuckets} does not contribute to query complexity, but contributes $O(hd)\in O(nd)$ to the running time, using radix sort. 
Everything considered, the query complexity and running time of the bucketing phase are $O(nd\;\log (d\hat{k}/\delta))$.\\

\noindent{\bf Phase (ii) Eliminating irrelevant buckets.} 
The buckets that are tested for emptiness are those of type (i), (ii) and (iv) because buckets of type (iii) are dropped at line~\ref{line:dropdominated}. 
The number of buckets of type (ii) is at most $k$.
Furthermore, a bucket can be of type (iv) iff there is one dimension $i$ such that they share the same coordinates $I_i$ as a skyline bucket on dimension $i$, and the interval $I_i$ is not trivial.
\bnote{In the literature~\cite{AfratiKSU12}, one finds the notion of relaxed skyline: the set of points that are not strictly dominated. The buckets we keep (ii) and (iv) form the relaxed skyline of buckets}\fnote{are you suggesting that we also use this notation? We are  not sure it would make the exposition much better, but we trust your judgement}\bnote{Not particularly needed, it was just a remark, a priori for us only, to compare with existing work.}
Consequently, by Lemma~\ref{lem:assignment-properties}, there are at most $(dk)\cdot 4n/(d\hat{k}^2)$ points that belong to buckets of type (iv). 
The number of points in buckets of type (ii) is even smaller: when such a bucket is trivial it contains only skyline points, and when it is not trivial, there is a dimension on which it is a non-trivial interval and therefore by Lemma~\ref{lem:assignment-properties} it has at most $4n/(d\hat{k}^2)$ points, hence a total of at most $k\cdot 4n/(d\hat{k}^2)$ points in buckets of type (ii).
When the estimate $\hat{k}$ is large enough ($k\in O(\hat{k})$), the number of points in buckets of type (ii) or (iv) is therefore $O(n/\hat{k})$. 
The case when this is not $O(n/\hat{k})$ because the estimate is not large enough is handled on line~\ref{line:check-cardinality-survivors}.
Similarly, Lemma~\ref{lem:assignment-properties} guarantees that $O(n/\hat{k})$ points have been assigned to buckets of the first kind. 
Therefore, the total number of points ever considered on line~\ref{line:empty} is $O(n/\hat{k})$. 
The contribution of line~\ref{line:empty} to the complexity is therefore $O(d(n/\hat{k})\log(\hat{k}/\delta'))$ by Lemma~\ref{lem:emptiness-test}.
Line~\ref{line:dropdominated} does not contribute to query complexity, but contributes $hd\hat{k}\in O(nd\cdot\min(\hat{k},|\sky(X)|))$ to the running time. 

Actually, we need to optimize a bit the algorithm to achieve that running time. 
There can be much more than $\hat{k}$ iterations, but there are only $\min(\hat{k},|\sky(X)|)$ "relevant" iterations in which we need to drop buckets.
So we first strengthen the requirement on the order at line~10, so that a bucket comes before buckets it \emph{weakly} dominates, where $B'$ weakly dominates $B$ (using the notation above) if in every dimension $\max_i\leq \max I_i'$. 
At line~17, if $B_i$ has already been marked as weakly dominated, we move on to the next iteration (any bucket that $B_i$ would dominate has already been dropped). Otherwise, we iterate through the list of remaining buckets, and we perform the following operations at a cost of $O(d)$ per bucket: we drop the buckets that $B_i$ dominates, and mark the other buckets that $B_i$ weakly dominates. There are only $\min(\hat{k},|\sky(X)|)$ buckets that are not weakly dominated, hence the running time.
\bnote{Actually our algorithm is currently checking emptiness on weakly dominated buckets. But we could alternatively just add all weakly dominated buckets to reduced pb, including the ones that only contain incorrect assignments, and possibly even including the ones that are strictly dominated by some other bucket. This could be performed at line~17: a same iteration through all buckets would both drop dominated buckets (from the remaining buckets at least, and possibly from the ones that we planned to send to the reduced problem), and add the other weakly dominated buckets to the reduced problem. Asymptotically, worst-case number of comparison and size of reduced set are similar because there are not too many erroneous assignments. Running time would be the same.}\\

\noindent{\bf Phase (iii) Solving the reduced problem.} 
Finally, at line~\ref{line:GM} the size of $X'$ is $O(n/\hat{k})$, so its skyline can be computed in $O(nd\log(\hat{k}/\delta'))$ by $\guessb$.\\

We next show that the correct answer is returned with high probability. 
First, the probability that the algorithm fails to satisfy our requirements (a) to (c) above are respectively $d\cdot \delta'/d$, $\delta/\hat{k}$ and $\hat{k}\cdot \delta'/\hat{k}+\delta'$. So the conditions are met --- hence the algorithm returns the correct output --- with probability at least $1-4\delta$.
\end{proof}

\subsection*{Algorithm for Theorem~\ref{thm:highdim}}
\begin{algorithm}[H]
\caption*{$\operatorname{\bf Algorithm}$\ $\guess(X,\delta )$     \  \ \ \ \ (see Theorem~\ref{thm-lowdim})
\\
{\bf input:} $X$ set of points, $\delta$ error probability
\\
{\bf output:} $\sky(X)$\\
{\bf error probability:} $\delta$
}
\begin{algorithmic}[1]
\label{alg1}
\STATE$\hat{k}\leftarrow (\lfloor d/\delta\rfloor )^2$ 
\REPEAT 
	\STATE  $\delta\leftarrow \delta/2$ ; $\hat{k}\gets \hat{k}^2$ ; $S\leftarrow$ $\alga( \hat{k} ,X,\delta)$ 
\UNTIL{ $|S| < \hat{k} $} 
\STATE Output $S$
\end{algorithmic}
\end{algorithm}

\section*{Appendix: Lower Bounds}

It is relatively straightforward to prove that any algorithm computing skylines with error probability ar most $\delta$  requires $\Omega(dn\log (1/\delta))$ expected queries (\cite{GM15},\cite{FRPU94}), for arbitrary $d$ and $n$. 
In this section, we thus fix $\delta$ to $1/10$ and aim to establish a $\Omega (dn\log k)$ lower bound on the query complexity. 
Together, those bounds show that
computing skylines requires $\Omega(dn\log(k/\delta))$ comparisons, as claimed in Theorem~\ref{THM:LOWER}. 

We denote by $\model$ the following problem: 
the  input is a set $X$ of $n$ points with dimension $d$, whose skyline has exactly $k$ skyline points. The goal is to return the skyline with error probability at most $1/10$. We assume that input comparisons have error probability at least  $1/3$  (and at most $2/5$), and assume w.l.o.g. that $k>10$.
To prove that any algorithm solving $\model$ must use $\Omega(nd\;\log k)$ queries in expectation, 
we define a noisy vector problem, 
in which one is given $h$ vectors each of length $\ell$ and needs to decide for each vector whether it is the all-zero vector. 
We prove a lower bound for this noisy vector problem, and reduce it to $\model$ to obtain our lower bound on $\model$.

 \subsection{$(h,\ell)$-\segproblem: Definition and Lower Bound }

In the $(h,\ell)$-\segproblem
the input $S$ is a collection $\{\mathbf{v}^1, \mathbf{v}^2,\dots,\mathbf{v}^h\}\subseteq \{ 0,2\}^{\ell}$ of vectors 
such that for each $i\in [h]$, $\sum_{j=1}^{\ell} \mathbf{v}_j^i\leq 2$, and the output
is a vector $(w_1,w_2, \dots, w_h) \in \{0,2 \}^h$ such that for each $i\in [h]$, $w_i=\sum_{j=1}^{\ell} \mathbf{v}_{j}^i$.
We define the distribution $\mu$  over vectors of $\{ 0,2\}^{\ell}$ as follows. For each $j\in [\ell]$, $\mu(2e_j)=1/(2\ell)$, where $e_j$ is the canonical vector with a $1$ in the $j$-th entry and zero elsewhere; $\mu(0,\ldots,0)=1/2$.
For inputs to $(h,\ell)$-\segproblem, we will consider the product distribution $\mu^k$.

\begin{lemma}
\label{pferd}
For $(h,\ell)$-\segproblem under the product distribution $\mu^h$, if $A$ is a deterministic algorithm with success probability at least $3/4$, then the worst case number of queries of $A$ is $\Omega(\ell h \log h)$.
As a consequence, any (possibly randomized) algorithm solving the $(h,\ell)$-\segproblem with success probability $3/4$ requires at least $\Omega(\ell h \log h)$ queries (on a worst-case input).
\end{lemma}
\begin{proof}
The proof is by contradiction. Assume that $A$ is an algorithm with success probability at least $3/4$ and worst case number of queries $T\leq (\ell h \log_3 h )/1000$. 
We assume that the adversary is \emph{generous}, \ie the adversary tells the truth for every entry $(i,j)$ such that $v^i_j=0$, and lies with probability $1/3$ otherwise. 
Generalizing the two-phases computational model  by Feige, Peleg, Raghavan and Upfal~\cite{FRPU94}, we will give the algorithm more leeway and study a 4-phase computation model, defined as follows. In the first phase, the algorithm queries every entry $v^i_j$  $(\log_3 h)/100$ times. In the second phase, the adversary reveals to the algorithm all remaining hidden entries $(i,j)$ such that $v^i_j=2$, except for a single random one.  In the third phase, the algorithm can strategically and adaptively choose $h \ell/10$ entries, and the adversary reveals their true value at no additional cost. Finally,  in phase 4, the algorithm outputs $w_i=2$ for every vector where it found an entry equal to $2$, and $w_i=0$ for the rest of the vectors.

To see how the two models are related, observe that since $T\leq (\ell h \log_3 h )/20$, by Markov's inequality at most a set $S$ of $\ell h /10$ entries are queried by algorithm $A$ more than $(\log_3 h)/2$ times, so at the end of the first phase we have queried every entry at least as many times as $A$, except for those $\ell h/10$ entries, and in the beginning of the third phase there is all the necessary information to simulate the execution of $A$, adaptively finding $S$ (and getting those values correctly), hence the success probability of the three-phase algorithm is greater than or equal to the success probability of $A$. Also observe that, thanks to the definition of $\mu$ and to the generosity of the adversary, any execution where all queries to a vector lead to $0$ answers must lead to an output where $w_i=0$---else the algorithm would be incorrect when $\mu$ selects the null vector.

We now sketch the analysis of the success probability of the three-phase algorithm. Due to the definition of $\mu$, with probability at least $9/10$ the ground-truth input drawn from $\mu^h$ has $h/2\pm O(\sqrt{h})$ vectors that contain an entry equal to $2$. At the end of the first phase, and due the fact that the adversary is generous, we have that most of them have been identified. There remain  $h/2\pm O(\sqrt{h})$ vectors that appear  to be all zeroes, and about $(h/2)(1/3)^{(\log_3 h)/2}=(1/2)\sqrt{h}$ of those vectors contain a still-hidden entry whose true value is 2. During the second phase, all of those hidden 2's are revealed except for one. At that point, there still remain $h/2\pm O(\sqrt{h})$ vectors whose entries appear to be all zeroes, there is a 2 hidden somewhere uniformly at random, but all entries have been queried an equal number of times, all in vain. To find that remaining hidden entry (and therefore decide which $w_i$ is equal to 2), the algorithm has no information to distinguish between the $\ell (h/2\pm O(\sqrt{h}))$ remaining entries.
Since, the algorithm may only select $\ell h/10$ elements to query further, the algorithm's success probability after the fourth phase cannot be better than $(h\ell/10)/(\ell (h/2\pm O(\sqrt{h})))<1/4$, a contradiction.

The first claim of the Lemma implies the second by Yao's principle.
\end{proof}

\subsection{Reduction: Proof of Theorem~\ref{THM:LOWER}} 
\label{sec:reduction}

Our reduction will prove the lower bound $\Omega(dn\log k)$ for any $n,d,k>0$ such that $d-2$ divides $k$, $n\geq k^5$ and $n$  is of the form $n=(\ell+d-2)k/(d-2)$ for some $\ell>0$.\\

\begin{figure}[H]
\begin{center}
\includegraphics{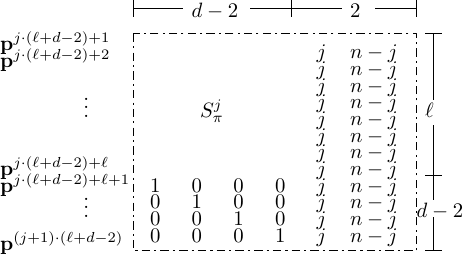}	
\end{center}
\caption{Block $(j,n-j)$ of the reduction. The vectors of $S^j_\pi$ placed in this block are $\mathbf{v}^{ j(d-2)+1}, \mathbf{v}^{ j(d-2)+2}, \dots, \mathbf{v}^{ j(d-2)+(d-2)}$.
 }
\label{fig:reduction}
\end{figure}

\noindent {\it Step 1.} 
Let $k, \ell>0$.  Let $d>0$ such that $d-2$ divides $k$. Let $n=(\ell+d-2)k/(d-2)$. We assume $n\geq k^5$.
From an input $S=\{ \mathbf{u}^1, \mathbf{u}^2,\dots,\mathbf{u}^k\}$ to the $(k,\ell)$-\segproblem,
we first show how to construct an input $\mathcal{I}_S$ for $\model$ with $n$ points in $d$ dimensions and a skyline that is likely to be of size $k$, where $n=(\ell+d-2)k/(d-2)$. We first randomly permute the entries of each $\mathbf{u}^i$, by using $k$ independent permutations, resulting in $S_\pi=\{\mathbf{v}^1, \mathbf{v}^2,\dots,\mathbf{v}^k\}$.
Partition $S_\pi$ into $k/(d-2)$ blocks of $d-2$ vectors, where for  $j \in \{0,1, \dots, k/(d-2)-1 \}$, block 
$S^j_\pi = \{ \mathbf{v}^{ j (d-2) + i} \colon  i \in [d-2] \} .$
For each block, define $\ell+d-2$ points, as displayed (one point per row) on Figure~\ref{fig:reduction}, and the union over all blocks  is the input $\mathcal{I}_S$ to the $\model$. Formally, we define point $\mathbf{p}^{(t)}$ with $t= j \ell + i$ as follows.

\[ 
\mathbf{p}^{(t)}_{r} :=
\begin{cases}
	j & \text{ if $r=d-1$},\\
	n-j & \text{ if $r=d$},\\
	1 & \text{ if $r=i-\ell$ and $\ell \leq i $ },\\
	 v^{j (d-2)+i}_{i} & \text{ if $r=i$ and $ i \in [d-2]$,}\\
		0 & \text{ otherwise.}\\
\end{cases}
\]

 \noindent {\it Step 2.} Because of the non-domination implied by the last two coordinates of any point, the skyline of the set of points is the union over all blocks of the skyline of each block. Fix an arbitrary block and focus on the first $d-2$ dimensions. For each dimension, the corresponding column (whose first $\ell$ coordinates are those of some vector $\mathbf{v}^i$) contains exactly one $1$ (on the row of some point $\mathbf{p}$) and possibly one $2$, the remaining entries being all $0$.
It is easy to verify that  $\mathbf{p}$ is  part of the skyline  if and only if $\mathbf{v}^i =\vec{0}$. \\

From the output $\sky(I)$
it is now easy to construct the output of the $(k,\ell)$-\segproblem:  
For all blocks, for all dimensions $\leq d-2$,  if $\mathbf{p}\in \sky(I)$  then $w_i\gets 0$ else $w_i\gets 2$.  
This yields the correct output $\mathbf{w}=(w_1, w_2, \dots, w_k)$.
 Thus we derive the following observation.
  \begin{observ}\label{obs1:numberofskylinepoints}	
 Given the set of points $\sky(\mathcal{I}_S)$,  one can recover the solution to the $(k,\ell)$-\segproblem without further queries.
 \end{observ}

Furthermore, in the following we prove that the construction is likely to have $k$ skyline points.
 \begin{lemma}\label{obs2:numberofskylinepoints}
Let $\mathcal{E}$ be the event that the input $\mathcal{I}_S$ has exactly $k$ skyline points. Then, $\Pr{\mathcal{E}} \geq 1-1/k$.  
 \end{lemma}
 \begin{proof}
 First observe that, by construction, regardless of whether $\mathcal{E}$ holds, every block contains at most $d-2$ skyline points: Consider an arbitrary block. The last two dimensions are identical for each point belonging to that block and we focus thus on the first $d-2$ dimensions. 
  There are exactly  $d-2$ points with one coordinate being $1$ and all of these points are potential skyline points. In particular, take any such point $\mathbf{p}$ and assume that the $i$'th coordinate of $\mathbf{p}$ is $1$. Then $\mathbf{p}$ is part of the skyline if and only if  the vector $\mathbf{v}^i$ is the null vector. 
 Moreover,  every block can have at most $d-2$ entries with value $2$ and each such $2$ eliminating one potential skyline point. Thus, there are at most $d-2$ skyline points per block. 

Consider the vectors $\mathbf{v}^{i_1}, \mathbf{v}^{i_1+1}, \dots, \mathbf{v}^{i_2}$ of any block.
We say they are \emph{collision free} if the following holds: if $\mathbf{v}^{j}_{j^*}=2$ for $j\in [i_1, i_2]$, then $\mathbf{v}^{j'}_{j^*}=0$ for all $j' \in [i_1, i_2] \setminus \{ j \}$. 
Observe that if the vertices of any block are collision free, then each of the first $d-2$ dimensions 
is dominated by a distinct skyline point and thus there $d-2$ skyline points in that block. Thus, if the vectors of every block are collision free, then
there $d-2$ skyline points per block and summing up over all $k/(d-2)$ blocks, we get that there are thus $k$ skyline points in total.  

Thus, in order to bound $\Pr{\mathcal{E}}$ it suffices to bound the probability that all blocks are collision free..
Recall that the random permutations $\pi_1, \pi_2, \dots, \pi_k$ permute each vector $\mathbf{v}^i$ independently. 
Since in a block at most $k^2$ pairs may collide, and each collision happens with probability $1/\ell$,  the expected number of \emph{collisions} per block is at most $k^2/\ell$.
 The expected number of collisions over all blocks is thus, by the union bound, at most $(k/(d-2))\cdot  k^2/\ell \leq 1/k$, by our assumption that $k^5\leq n$.
 Thus, the claim follows by applying Markov inequality.
 \end{proof}

\begin{proof}[Proof of Theorem~\ref{THM:LOWER}]
 	Suppose for the sake of contradiction that there exists an algorithm ${\A}$ recovering the skyline for any input with exactly $k$ skyline points, with error probability at most $1/10$, and using $o(nd\log k)$ queries in expectation. By Markov inequality, the probability that the number of queries exceeds 20 times the expectation is at most $1/20$, so truncating the execution at that point adds $1/20$ to the error probability, transforming $\A$ into an algorithm $B$ that recovers the skyline for any input with exactly $k$ skyline points, with error probability at most $1/20+1/10$, and using $o(20n d\;\log k)$ queries in the worst case.
 	We next show that this implies that one can solve the $(k,\ell)$-\segproblem with $o(n d\; \log k)$ w.p. at least $3/4$. 
 	
 	Let  $S$ be the input of the $(k,\ell)$-\segproblem. We cast $S$ as an input $\mathcal{I}_S$ of $B$ as described in Section~\ref{sec:reduction}.
 	By Lemma~\ref{obs2:numberofskylinepoints}, the event $\mathcal{E}$ holds w.p. at least $1-1/k$
 	and thus there are $k$ skyline points. If so, $B$ can then compute the skyline with $o(n d\;\log k)$ queries with error probability at most $1/20+1/10$.
 
 	By Union bound, the probability that $B$ errs or that $\mathcal{E}$ does not hold is at most $1/k+1/20+1/10\leq 1/4$.
 	Thus, by Observation~\ref{obs1:numberofskylinepoints}, one can obtain w.p. at least $3/4$ the solution to $(k,\ell)$-\segproblem
 	using $o(n d\; \log k)$ queries, which means $o(\ell k \;\log k)$ queries by definition of $n$ and $d$. 
 	This is a contradiction to Lemma~\ref{pferd}. 
 	To conclude our proof,
    we observe that the assumptions that $n\geq k^5$ and $d-2$ divides $k$ are unnecessary and can be replaced with $d\leq k$. This is because we can introduce dummy points such as copies of existing skyline points, and copy on the following dimensions the values from previous dimensions to show that taking a larger value for $d$ or $k$ only makes the skyline problem harder.
 	\end{proof}

\end{document}